\documentclass[journal]{IEEEtran}

\usepackage{amsmath,graphicx}
\usepackage{pdflscape}
\usepackage{tabularx}
\usepackage{lipsum}
\usepackage{xcolor}
\usepackage{amssymb}
\usepackage{multirow}
\usepackage{breqn}
\usepackage{afterpage,lipsum}
\usepackage{blindtext}
\usepackage[noend]{algpseudocode}
\usepackage[linesnumbered,lined,boxed, commentsnumbered]{algorithm2e}
\usepackage{blindtext}

\newcommand\blfootnote[1]{%
\begingroup
\renewcommand\thefootnote{}\footnote{#1}%
\addtocounter{footnote}{-1}%
\endgroup
}

\title{\LARGE \bf
Activity Recognition from Newborn Resuscitation Videos}

\author{\O yvind~Meinich-Bache, 
		Simon~Lennart~Austnes,
        Kjersti~Engan,~\IEEEmembership{Senior Member,~IEEE,}
        Ivar~Austvoll,~\IEEEmembership{Member,~IEEE,}        
        Trygve~Eftest\o l,~\IEEEmembership{Senior Member,~IEEE,}   
        Helge~Myklebust,
        Simeon~Kusulla, 
        Hussein~Kidanto and    
        Hege~Ersdal}  
       
\begin{document}

\maketitle
\thispagestyle{empty}
\pagestyle{empty}%

\begin{abstract}
\textit{Objective:}
%In 2017 the average global mortality rate for newborns was 18 deaths per 1000 live births. 
Birth asphyxia is one of the leading causes of neonatal deaths. A key for survival is performing immediate and continuous quality newborn resuscitation. A dataset of recorded signals during newborn resuscitation, including videos, has been collected in Haydom, Tanzania, and the aim is to analyze the treatment and its effect on the newborn outcome. An important  step is to generate timelines of  relevant resuscitation activities, including \textit{ventilation},  \textit{stimulation}, \textit{suction}, etc., during the resuscitation episodes.
\textit{Methods:}
We propose a two-step deep neural network system, ORAA-net, utilizing low-quality video recordings of  resuscitation episodes to do activity recognition during  newborn resuscitation. 
%The videos are noisy videos of poor quality and sometimes the activities are also heavily occluded. To handle this, 
The first step is to detect and track  relevant objects using Convolutional Neural Networks (CNN) and post-processing, and the second step is to analyze the proposed activity regions from step 1 to do activity recognition using 3D CNNs.
\textit{Results:}
The system recognized the activities  \textit{newborn uncovered}, \textit{stimulation}, \textit{ventilation} and  \textit{suction} with a mean precision of 77.67 \%, a mean recall of 77,64  \%, and a mean accuracy of 92.40 \%. 
%
%The activity \textit{uncovered} was recognized with a \textit{precision} of 87.75  \% and a \textit{recall} of 83.99 \%, the activity \textit{stimulation}  with a \textit{precision} of 78.79  \% and a \textit{recall} of 74.59 \%, the activity \textit{ventilation}  with a \textit{precision} of 87.30  \% and a \textit{recall} of 90.64 \% and the activity \textit{suction}  with a \textit{precision} of 56.85 \% and a \textit{recall} of 61.32 \%. 
Moreover, the accuracy of the estimated number of Health Care Providers (HCPs) present during the resuscitation episodes was 68.32 \%.
\textit{Conclusion:}
The results indicate that the proposed CNN-based two-step ORAA-net could be used for object detection and activity recognition in noisy low-quality newborn resuscitation videos.
\textit{Significance:} 
A thorough analysis of the effect the different resuscitation activities have on the newborn outcome could potentially allow us  to optimize  treatment guidelines, training, debriefing, and local quality improvement in newborn resuscitation.
\end{abstract}

\blfootnote{
This work is part of the Safer Births project which has received funding from: Laerdal Global Health, Laerdal Medical, University of Stavanger, Helse Stavanger HF, Haydom Lutheran Hospital, Laerdal Foundation, University of Oslo, University of Bergen, University of Dublin – Trinity College, Weill Cornell Medicine and Muhimbili National Hospital. The work was partly supported by the Research Council of Norway through the Global Health and Vaccination Programme (GLOBVAC) project number 228203.
}
\blfootnote{
\O, Meinich-Bache, S. L, Austnes K, Engan, I, Autsvoll and T, Eftest\o l is with the Dep. of Electrical Engineering and Computer Science, University of Stavanger, Norway.
H, Myklebust with Laerdal Medical, Norway.
S. Kusulla with the Research Institute, Haydom Lutheran Hospital, Manyara, Tanzania.
H, Kidanto with the School of Medicine, Aga Khan University, Dar es Salaam, Tanzania.
H, Ersdal with the Faculty of Health Sciences, University of Stavanger, Norway, and the
Department of Anesthesiology  and Intensive Care, Stavanger University Hospital, Norway.
}

\section{Introduction}
% SA/OMB\\
In 2017 the average global mortality rate for newborns was 18 deaths per 1000 live births \cite{neonatalmortality}. Low- and middle-income countries account for 99 \% of deaths  for neonates under four weeks of age \cite{neonatalmortality2}. Birth asphyxia is one of the leading causes of neonatal deaths, and mortality rates due to this complication have not seen the same rate of improvement as other common causes of newborn mortality \cite{improvementrate}. 
The immediate
presence of properly trained and equipped Health Care Providers (HCPs)  lessens these preventable
newborn deaths \cite{neonatalmortality3}. %In low- and middle-income countries the
The main therapeutic resuscitation activities for birth asphyxia in this setting comprise the following; positive pressure \emph{ventilations} using a Bag-Mask Resuscitator (BMR), \emph{stimulation}, \emph{suction} using a Suction Device (SD), and keeping the newborn warm using a blanket \cite{wyckoff2015part}.

The collaborative research and development project Safer Births\footnote{www.saferbirths.com} aims to establish new knowledge and develop new products to support HCPs in low resource countries with the purpose of saving more lives at birth. Since 2013 the project has been collecting various data during newborn resuscitation episodes at Haydom Lutheran Hospital in Tanzania. 
The acquired data, such as ECG,  flow during ventilation, videos, and the newborn outcome, can be used to gain critical insight into the effects of the different resuscitation activities, as well as facilitating ongoing training of HCPs, debriefing, and continuous quality improvement.
This could be achieved by creating activity timelines from the collected data and study them together with information on the condition of the newborn during resuscitation, found from the ECG, and the resuscitation outcome.

\begin{figure*}[h]
\centering
\includegraphics[width=\textwidth]{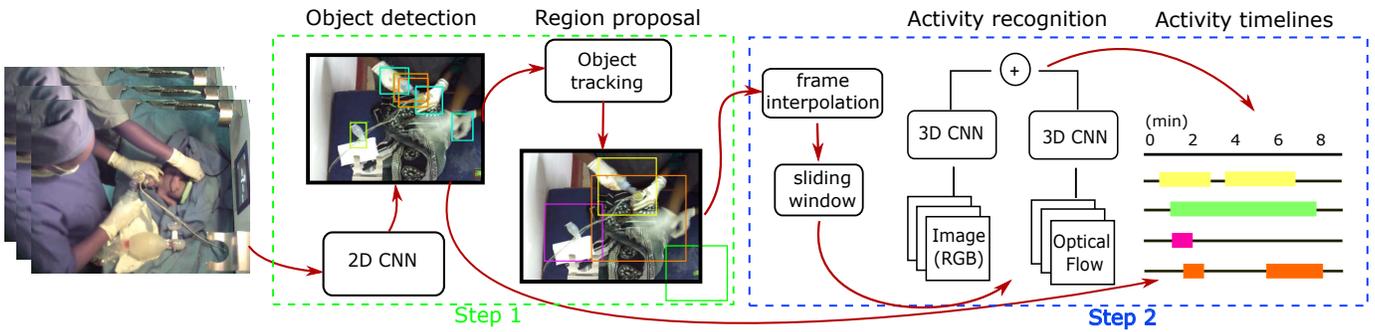}
\caption{Block scheme of the two-step ORAA-net for activity recognition from newborn resuscitation videos. Step 1: Object detection,  object tracking, and region proposal (presented in \cite{objectresuscitation}). Step 2: Activity recognition using 3D convolutional neural networks (CNNs) trained on the individual activities, and the generation of activity timelines. The \emph{frame interpolation} step in the activity recognition can be skipped for datasets of fixed and adequate frame rates.}  
\label{fig:oversiktPaper6}
\end{figure*}

An activity detector 
based on signals from a Heart Rate Sensor (HRS), (which is a prototype of the NeoBeat\footnote{https://laerdalglobalhealth.com/products/neobeat-newborn-heart-rate-meter/}), previously proposed by our research group in Vu et al., separated the activities \textit{stimulation}, \textit{chest compressions} and \textit{other} with a precision of 78.7 \%  \cite{huyen}. Automatic  analysis of video collected during newborn resuscitation could be utilized to potentially improve the precision achieved by using HRS signals, as well as to detect activities and information when the  HRS signals are not available. In addition, video analysis could also allow us to detect activities that are difficult or impossible to obtain from ECG and accelerometer measurements. 
%The video recordings available for analysis are of highly variable quality, recorded with different cameras of poor quality and with frame rates ranging from 0.5-30 frames per second (fps).

Video analysis of newborn resuscitation episodes has been documented to have a positive effect on both evaluation and training purposes \cite{skaare,gelbart, silvia, nadler}. However, such analysis involves manual inspection, which are both time-consuming and entails privacy issues. Hence, it would be beneficial to perform the analysis automatically.

Automatic video analysis and activity recognition using deep learning models has become very popular in the last few years. 
%Compared to recognizing spatial features in individual images, recognizing temporal information in an image series is far more complexed.
However, recognizing temporal information in an image series is a far more complex problem than recognizing spatial information in individual images.  Quite recently
DeepMind\footnote{https://deepmind.com/} and Carreira et. al \cite{i3d} proposed a two-stream activity recognition network, I3D, that utilized CNNs and transfer learning to achieve state-of-the-art results on the activity recognition dataset UCF-101. I3D is
based on  a 3D inflated version of the well-known CNN Inception v1 \cite{inception}, and Carreira et. al demonstrated that 3D CNNs can benefit from pre-trained 2D CNNs, and that transfer learning is highly efficient also in activity recognition. %The two streams used by I3D is RGB and optical flow estimated using the TV-L1 algorithm \cite{flow}. The models provided state-of-the-art results on the activity recognition dataset UCF-101.

In 2016 Guo et al.   proposed
an automatic activity detection system for newborn resuscitation
videos \cite{activityDetectionNewbornVideos}. The system was based on a pre-trained Faster RCNN
network and the \textit{person} class was used to detect the newborn and finding the region of interest. Further, linear Support-Vector Machines
(SVMs) were trained on individual video frames to perform  activity recognition. In our dataset the newborn is covered with a sheet most of the time, making a \textit{person} detection not the best approach. We have chosen a different approach that we consider more suited for our dataset and the activities we want to detect - which are not necessarily newborn position-dependent. Our approach for activity recognition is to learn deep neural networks  to recognize the typical movement for an activity by utilizing  sequential frames instead of individual frames in the activity analysis, e.g. the BMR used in \textit{ventilation} has to be in a correct position and  \textit{squeezed} in order to be assigned to the activity class. This approach is also more suited for our low-quality videos where it can be difficult to detect activities from individual frames due to motion blurring.
%The proposed system is called the ORAA-net and has two main %steps: 1) detect activity-relevant objects and track their %position, and 2) analyze the proposed object regions over %time to recognize activities that possibly overlap in time.
%
%
The proposed system consist of the main parts; Object detection, Region proposal, Activity recognition, Activity timelines, and is named ORAA-net for short.   We consider it as being a two-step approach where the first step comprise the OR, i.e detect and track relevant objects and propose regions surrounding them, and the second step comprise the AA i.e activity recognition and the generation of the activity timelines.

Results from the first step in the ORAA-net was presented in \cite{objectresuscitation}. The step utilized the YOLOv3 \cite{yolov3} object detection architecture and subsequent post-processing to propose regions surrounding the objects throughout the resuscitation videos. The performance of the object region proposal during activities was
97 \% on ventilation, 100 \% on attaching or removing the
heart rate sensor, and 75 \% on suctioning. 
%
%This step covered object detection and tracking performed on the collected video recordings by use of the YOLOv3 \cite{yolov3} object detection architecture and subsequent post-processing.  
%The performance of the object region proposal, based on the object detection and post-processing, during activities was
%97 \% on ventilation, 100 \% on attaching or removing the
%heart rate sensor, and 75 \% on suctioning. 
Additionally, the
number of HCPs present in the image frames
were estimated with a performance of 71 \%. Potential for
improvement for object detection, particularly in the detection of the suctioning device, was however recognized.

In this paper, we present results from step two of the  ORAA-net. Short sequences from the proposed regions are used as input to I3D models trained to recognize the different resuscitation activities and to generate activity timelines. 
%to train I3D models to recognize activities from the proposed regions, and to create timelines of the different resuscitation activities. 
Besides, the paper also presents improvements of the ORAA-net`s first step, which has been attained through experiments with three additional state-of-the-art object detection networks, and by proposing a method for finding the region surrounding the newborn.

The paper uses several acronyms and the most commonly used are listed below for increased readability.

\begin{table}[h]
\centering
%\caption{Acronyms}
\begin{tabular}{|l|l|}
\hline
\multicolumn{1}{|c|}{\textbf{Term}} & \multicolumn{1}{c|}{\textbf{Acronym}} \\ \hline
Heart Rate Sensor                   & HRS                                   \\ \hline
Bag-Mask Resuscitator               & BMR                                   \\ \hline
Suction Device                      & SD                                    \\ \hline
Health Care Provider                & HCP                                   \\ \hline
Health Care Provider Hand           & HCPH                                  \\ \hline
Inception 3D          & I3D                                  \\ \hline
Linear Frame Interpolation          & LFI                                  \\ \hline

\end{tabular}
\end{table}
%
%The results obtained with the Inception 3D (I3D) architecture \cite{i3d} - a 3D inflation of the image classifier Inception v1 \cite{inception} - trained further with both optical flow and RGB representations of the resuscitation activities.
%Due to the highly variable frame rate, linear frame interpolation was applied to the videos prior to the generation of training data  to create videos with an adequate high and fixed frame rate. %  and the activity recognition. 
%Besides, the paper also presents an improvement in the object detection of the system`s first step, which has been attained through experiments with three additional state-of-the-art object detection networks. For recognition of activities that do not depend on objects, such as \textit{is the newborn covered or not}, we propose a method for localization of the newborn - that is the Region of Interest (ROI) - based on the detected HCPs` hands. 
%Figure \ref{fig:oversiktPaper6} shows the total proposed system for the generation of the activity timelines.

\label{sec:intro}

\begin{table*}[h]
\centering
\caption{Significant architectural features of the object detection networks.  * Base feature extractor proposed in the original design.} 

\begin{tabular}{|l|c|c|c|c|}
\hline
& \textbf{YOLOv3} \cite{yolov3} & \textbf{RetinaNet} \cite{retinanet} & \textbf{SSD MultiBox} \cite{ssdmultibox}& \textbf{Faster R-CNN} \cite{frcnn}\\ 
%& \cite{yolov3} &  \cite{retinanet} & \cite{ssdmultibox} & \cite{frcnn} \\ 
\hline
\textbf{Base CNN} & Darknet53* & Optional (ResNet-50) & VGG-16* & Optional (ResNet-50) \\ \hline
\textbf{Approach} & One-stage & One-stage & One-stage & Two-stage \\ \hline
\textbf{Feature pyramid network} & Yes & Yes & No & No \\ \hline
\textbf{\# Feature map scales} & 3 & 5 & 6 & 1 \\ \hline
\textbf{Anchors} & 9 & 9 & 6 & 9 \\ \hline
\textbf{Hard negative mining} & No & No & Yes & No \\ \hline
\textbf{Classification loss function} &  \begin{tabular}[c]{@{}c@{}} Binary       cross-entropy  \end{tabular} & Focal loss & \begin{tabular}[c]{@{}c@{}} Categorical      cross-entropy  \end{tabular}  & \begin{tabular}[c]{@{}c@{}} Categorical     cross-entropy  \end{tabular} \\ \hline
\textbf{Regression loss function} & \begin{tabular}[c]{@{}c@{}} Sum of      squared errors  \end{tabular}  & Smooth L1 & Smooth L1 & Smooth L1 \\ \hline

\end{tabular}
\label{tab:table1Paper6}
\end{table*}

\section{Objectives}
We aimed to recognize the following therapeutic activities, provided in this setting and known to affect the condition of a newborn during resuscitation  \cite{wyckoff2015part}:
\begin{itemize}

\item \emph{Uncovered} - the newborn is not covered by a blanket.
\item \emph{Stimulation} - thoroughly drying and rubbing the newborn.
\item \emph{Ventilation} - positive pressure ventilation using a BMR.
\item \emph{Suction}: removal of liquid from the mouth/airways using a SD, where in this datamaterial the \textit{Penguin}\footnote{https://laerdalglobalhealth.com/products/penguin-newborn-suction/} is used.
\end{itemize} 
In addition, we also aim to recognize other activities and parameters that could be of interest:

\begin{itemize}
\item \emph{Attaching/adjusting HRS}  - relevant for the analysis of the ECG signals collected by the HRS.
\item \emph{Remove HRS} - relevant for the analysis of the ECG signals collected by the HRS.
\item  \emph{Number of HCPs} treating the newborn - might have an impact on the newborn outcome. 
\end{itemize}

\emph{Uncovered} and \emph{Stimulation} could be detected by analyzing an area around the newborn, %, found from the position of the detected HCPH during the resuscitation episode. 
\emph{Ventilation}, \emph{Suction}, \emph{Attaching/adjusting HRS}, and \emph{Remove HRS} by tracking the objects BMR, SD, and HRS, and by analyzing their surrounding areas, and finally, \emph{Number of HCPs} by counting the number of detected HCPH.

\section{Data material}
\label{sec:data}

The dataset was collected at Haydom Lutheran Hospital in Tanzania using Laerdal Newborn Resuscitation
Monitor (LNRM) \cite{huyen} and with cameras mounted over the 
resuscitation tables. The dataset contains 481 newborn resuscitation episodes with  
video, LNRM data, state of the newborn during resuscitation, and information on the newborn outcome. The videos only include the time period when a newborn is placed on the resuscitation table and the duration of the videos range from 1-60 minutes with a median duration of approximately 7 minutes. %In addition, from the ECG signals in the LNRM data we can also extract information on the state of the newborn during the resuscitation.
In this work, 96 randomly selected videos from the dataset were used to develop and evaluate the performance of the proposed system. The LNRM signals were recorded by  measuring signals with a BMR and a HRS, both connected to the LNRM. The HRS of the LNRM is an early version of the \textit{NeoBeat\footnote{https://laerdalglobalhealth.com/products/neobeat-newborn-heart-rate-meter/}}. The recorded videos were not initially intended for automatic video analysis, but rather as support material for human interpretation when needed. As a consequence, no standardization in camera type and camera settings were applied. The videos are recorded with different kinds of low-quality cameras and have variable frame rates - ranging from 0.5-30 fps, resolutions, focus settings  and quality. Furthermore, there are also variations in the  position of the mounted cameras and in light settings in the labor rooms. These variations make it more challenging to perform object detection and activity recognition.

\section{Methods}
\label{sec:methods}

An overview of the proposed ORAA-net is illustrated in Figure \ref{fig:oversiktPaper6}. The system is divided into 2 main steps - 1 - object detection and region proposal, and 2 - activity recognition and timeline generation,  and they are explained seperately in the following.  %Both the method section and the experimental section is divided into part A, B, and C corresponding to the naming in Figure \ref{fig:oversiktPaper6}.  The method is explained in the following. 

%Part A.1 and A.2 in the figure, we have compared state of the art object detectors and used the best one to perform predictions on the newborn resuscitation videos. In section B we have post processed the detections in the same way as explained in \cite{objectresuscitation} to perform object tracking and region proposal. In this step we have also now added a method for localizaation of the newborn region based on the detected HCPs hands. In section C- Activity detection, step 4, we have trained 3D inflated convolutional networks, Inception 3D \cite{i3d}, to perform activity recognition on each of the different resuscitation activities. In step 5 we have the final output of the system, the activity timelines.

\subsection{ORAA-net Step 1 - Object Detection and Region Proposal}
\label{sec:objectDetectionPaper6}

%Convolutional Neural Networks (CNNs) have, in recent years been used to attain record-breaking results in the prominent object classification, segmentation, and detection challenges like the PASCAL Visual Object Challenge \cite{pascalvoc}, Common Objects in Context \cite{coco} and ImageNet Large Scale Visual Recognition Challenge \cite{imagenet}. 
%Some NNs  classify the entire image as belonging to one class, in principle classifying the dominant object in an image.  Object detection networks, however, are tasked with classifying and locating \emph{all} objects, from a set of interest objects, in each image.  
%These networks are divided into two types of classes, one-stage or two-stage, based on their approach to selecting candidate object regions from images. Two-stage networks runs a region-proposal step before detection, while one-stage networks runs detection over a dense collection of candidate regions. Large numbers of candidate regions presents an class-imbalance issue of objects vs. background training examples. One-stage detectors lessens the class-imbalance issue through different methods, such as RetinaNet's Focal Loss \cite{retinanet} or Hard Negative Mining (HNM) used in SSD MultiBox \cite{ssdmultibox}. Some networks produce detections from feature maps of multiple scales to improve scale-invariance. These feature maps are often output from a Feature Pyramid Network (FPN) \cite{pyramid} built into the architecture. 

In our previous work we achieved encouraging results using the YOLOv3 architecture as the object detector on the presented dataset and challenge \cite{objectresuscitation}.  However, especially the small SD, (labled SP in \cite{objectresuscitation}) had improvement potential.  In this work we implement, further trained and tested RetinaNet \cite{retinanet}, SSD MulitBOx \cite{ssdmultibox} Faster R-CNN  \cite{frcnn}, in addition to YOLOv3 \cite{yolov3} on our dataset to find the best solution for step 1 (see Figure \ref{fig:oversiktPaper6}). A comparison of the main features of the object detection networks considered in this work  are shown in Table \ref{tab:table1Paper6}. 
%
%A comparison of the main features of the object detection networks is shown in table \ref{tab:table1Paper6}.  

% \begin{table}[h] % TRANSPONERT VERSJON
% \centering
% \caption{Comparison of significant architectural features of the object detection networks.} 

% \begin{tabular}{|l|c|c|c|c|c|c|c|c|}
% \hline
% & Base & Approach & \# Feature & Anchors & Classification & Regression & NMS & HNM \\
%  & CNN & & map scales & & loss function & loss function & & \\
% \hline
% \textbf{YOLOv3} \cite{yolov3} & & & & & & & & \\ \hline
% \textbf{RetinaNet} \cite{retinanet} & & & & & & & & \\ \hline
% \textbf{SSD MultiBox} \cite{ssdmultibox} & & & & & & & & \\ \hline
% \textbf{Faster R-CNN} \cite{frcnn} & & & & & & & & \\ \hline
% \end{tabular}
% \label{tab:table1}
% \end{table}

%\subsection{Region Proposal - Object Tracking - Methods}
\label{sec:regionProposalPaper6}

Object tracking and region proposal of the class BMR, SD and HRS are performed on the object detection results as follows: %described in detail in \cite{objectresuscitation}. In short: 

\begin{itemize}
  \item Localize the most likely true object position in each image using the object detections probability scores.
  \item Fill detection gaps by choosing the previous detected value.  
    \item Remove short peaks by checking, in time, if a rapid position change is an actual large position change or if the position quickly returns to the same area as prior to the change.
  \item Signal smoothing using a moving average filter.
    \item Region proposal for further activity analysis. 500 $\times$ 500 pixel regions around the tracked objects as shown in  Figure \ref{fig:oversiktPaper6}, step 1.

\end{itemize}
This is consistent with the method we proposed in \cite{objectresuscitation}, were more details can be found. 
%The timeline estimation of the number of HCP present in the resuscitation episode, $\mathit{\#HCP(i)_{E}}$, is found by counting the number of detected health care providers hands (HCPH) for each time index $i$. % as explained in \cite{objectresuscitation}.

In this work, we propose an additional region of interest to further analyze; the newborn region.
%In addition to the proposed regions around the objects BMR, SD and HRS we also determined the most likely newborn position. 
Analyzing this region would make it possible to detect the activities which are not object dependent, like if the newborn is covered or not. Moreover, the newborn region may also allow us to recognize object dependent activities  for cases where the object tracking is poor.

\begin{figure}[h]
\centering
\includegraphics[width=\linewidth]{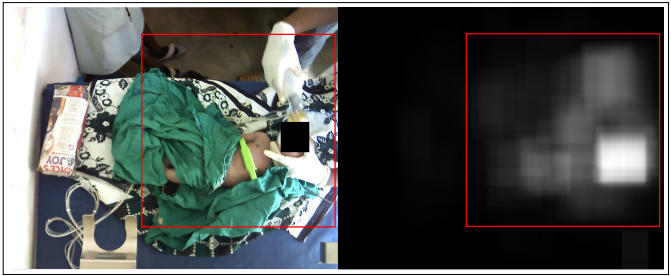}
\caption{Left: Example frame from a video. Right: Heatmap generated from the positions of the health care provider hands (HCPH) during the video. The red square illustrate the fixed 700 $\times$ 700 pixel sized newborn detection region.}  
\label{fig:heatmap}
\end{figure}

For each episode, a fixed newborn region is  found by first generating a heatmap of the whole image, $\mathit{HM(x,y)}$, where $x$ and $y$ are pixel coordinates. % and $E$ the resuscitation episode. 
The $\mathit{HM(x,y)}$ are initialized with zeros, and for each detection of a HCPH a value of 1 is added to the pixel area of the detection.
%
%
%In the generation of  $\mathit{HM(x,y)_{E}}$
%Further, 
Denote $\mathit{pA_{i,HCPH(n)} = \{x^{i,HCPH}_n, y^{i,HCPH}_n\}}$ to represent the pixel area of each detected HCPH, $n$, of the total HCPH detections, $N_{i}$, in frame $i$:

$\forall$ $i$ and \emph{For} $n=1:N_{i}$ \emph{do}:
\begin{equation}
\mathit{HM(x,y)} = \left\{ 
  \begin{array}{l l}
    \mathit{HM(\cdot)} + 1,\\ 
     \quad \quad \forall \{x,y\} \in  \mathit{pA_{i,HCPH(n)}} \\
    \mathit{HM(\cdot)},  \quad  \text{otherwise}
  \end{array}
\right .
\label{eq:diffIm}
\end{equation}
%The fixed 700 $\times$ 700 sized newborn region is selected by  finding the $x_m$ and $y_m$ that
%
The fixed and squared newborn region with size $R_s = 700$ pixels, is selected by  finding the $x_m$ and $y_m$ that

\begin{equation}
 \{x_m, y_m\} = argmax_{x_j,y_k} \sum_{m=x_j}^{{x_j+R_s - 1}} \sum_{n=y_k}^{{y_k+R_s - 1}}HM(m,n)
\end{equation}
%
%
%that maximize $\mathit{sum(HM(x_j:x_j+699,y_k:y_k+699)_{E,i}})$, 
where $\mathit{x_j \in \{1:im_{width}- (R_s -1) \}}$ and $\mathit{y_k \in \{1:im_{height}-(R_s - 1)} \}$. %,   is proposed as the fixed newborn region. 
An example of the generated heatmap with its proposed region is shown in Figure \ref{fig:heatmap}.

\subsection{ORAA-net Step 2 - Activity Recognition and Activity Timelines}

\subsubsection*{Dataset pre-processing}
\label{dataGenPaper6}

\begin{figure}[h]
\centering
% \hspace*{-1cm} 
\includegraphics[width=250px]{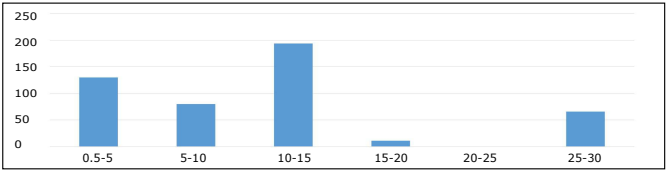}
\caption{Average video frame rates for the 481 videos in the dataset. X-axis is the video frame rate groups with frame rate interval of five, and Y-axis is the number of videos. }  
\label{fig:distributionPaper6}
\end{figure}

An important  step in  
activity recognition is to ensure that the data is of sufficient quality. This is especially important in our case where the video frame rates range from 0.5-30 fps.
% 
% 
%  sampled with a sample rate that are sufficiently high for recognition of the relevant activities. 
For videos with a very low frame rate it is difficult to separate the repetitive activities we are searching for,
%
%Videos with a very low frame rates make it difficult to separate our repetitive activities, 
such as \textit{stimulation}, where the HCP typically rubs the baby`s back,   from  random movements. 
We have  observed that for frame rates below 5 fps it can be very difficult to identify stimulations even by careful visual inspection. Thus, only videos with frame rates $>$ 5 fps are included in the dataset for training.  Videos with frame rates of 5 fps or lower accounts for 27 \% of the original dataset and the distribution of average fps for all videos can be seen in Figure \ref{fig:distributionPaper6}.

%
%
%Thus, some pre-processing steps are performed, and the first step is to remove videos below 5 fps from the training data. % are excluded from the generation of training data. 
%For videos below this threshold it is sometimes very difficult to separate a repetitive HCPH movement from a short $stimulation$ sequence by visual video inspection.
%
Thereafter, a pre-processing step is performed to convert the videos, now ranging from 5-30 fps, to a fixed and adequate frame rate. % by Linear Frame Interpolation (LFI) \cite{LFI}. % to generate videos of a fixed frame rate. 
%Let $f(k)$ represent the original video, $If(j)$ the LFI video
%and $ti$ the desired time for a specific frame in $If(k)$. The image frame for $ti$ is generated by: 
%
Although videos with frame rate below 5 fps are now removed, many of the remaining videos are still of low quality. Thus,  advanced up-sampling techniques that include motion analysis and require a certain frame rate, would not be well-suited, and  a simple  Linear Frame Interpolation (LFI) \cite{LFI} technique is chosen for the up-sampling. 
%
%Due to the poor image quality and the low frame rates, the videos lacks to much information information is already missing in the videos. Thus, a more complex video interpolation method that utilize 
%
%
% a simple Linear Frame Interpolation (LFI) technique is chosen for the video up-sampling. 
%LTI gives shadow effects and motion blurred images when large motions occur, but such effects are also present in original videos with sufficient frame rates. %, thus a more complex method that utilize motion compensation is
The artifacts from the LFI have a visual appearance similar to the blurring in some of the videos.
To represent the implementation of the LFI technique, first let f(t) be a frame at time $t$ from the original video. Given frames at times $t_1$ and $t_2$ we construct a new frame for time $t_i$ ($t_1<t_i<t_2$) by:

%
% Especially in this dataset where  videos 
%%
%has frame rates ranging from 2-30 frames per second (FPS) and some videos are of very poor quality.
% 
% to convert the videos to a fixed and adequate frame rate. %XX et al. \ref{lowFPS}  demonstrated that the results could be greatly impacted by using videos with low frame rates. 
%Videos with a very low sample rates would make it difficult to separate our repetitive activities, such as \textit{stimulation}, from  random movements. Especially in this dataset where  videos 
%%
%has frame rates ranging from 2-30 frames per second (FPS) and some videos are of very poor quality.
%%
%%
%%are of very poor quality.  Since the newborn resuscitation videos has frame rates ranging from 2-30 frames per second (FPS) 
%%

\begin{equation}
f(ti) = c_1 \cdot f_{t_1} + c_2 \cdot f_{t_2}
\end{equation}

where 
\begin{equation}
c_1 = \frac{\delta t_1}{T_{12}}, \quad c_2 = \frac{\delta t_2}{T_{12}},
\end{equation}
and where $\delta t_1 = t_i-t_1$, $\delta t_2 = t_2-t_i$ and $T_{12} = t_2-t_1$.
%where  $ f_{t1}$ and $f_{t2}$ are the frames in  $f(k)$ that minimize the time distance around time $ti$, and where

%\begin{equation}
%w_A = \frac{ti-t1}{t2-t1}   ,\quad \quad   w_B = \frac{t2-ti}{t2-t1} 
%\end{equation}

%Five fps is an absolute low sample rate for being able to distinguish the activity \textit{stimulation} from random HCPH movements.
%
%  in a short video example the most critical activity in terms of activity sample rate of the activity, stimulation.

\subsubsection*{Activity Recognition}

%DeepMind \footnote{https://deepmind.com/} and Carreira et. al \cite{i3d} have quite recently proposed a two-stream activity recognition network, I3D, based on  a 3D inflated version of the well-known CNN Inception v1 \cite{inception}. Carreira et. al demonstrated that 3D CNN can benefit from pre-trained 2D CNN, and that transfer learning is highly efficient also in activity recognition. The two streams used by I3D is RGB and optical flow estimated using the TV-L1 algorithm \cite{flow}. The models provided state-of-the-art results on the activity recognition dataset UCF-101, and recently 
For the activity recognition in step two of the ORAA-net, we have chosen to use multiple versions of the Inception 3D (I3D) architecture  proposed by Carreira et. al \cite{i3d}. 
The I3D is a state-of-the-art activity recognition network that currently holds the first place on the UCF-101 Leaderbord\footnote{https://paperswithcode.com/sota/action-recognition-in-videos-on-ucf101}.
I3D utilize both RGB and Flow images during predictions and combining these two data representations have been proven by others to be important for solving activity recognition tasks \cite{i3d, flowTest, flow2}.
Moreover, the authors have made their pre-trained models publicly available\footnote{https://github.com/deepmind/kinetics-i3d} which allows us to benefit from transfer learning.
We have further trained these models on newborn resuscitation activities data to perform activity recognition on the proposed regions from Section \ref{sec:regionProposalPaper6} as shown in step 2 Figure \ref{fig:oversiktPaper6}.

%
%The I3D is a state-of-the-art network that currently holds the first place in the UCF-101 Leaderbord \footnote{https://paperswithcode.com/sota/action-recognition-in-videos-on-ucf101}.
%Moreover, the authors have made their pre-trained datasets publicly available, and the I3D utilize both RGB and Flow images during predictions, which have been proven by many to be important for solving activity recognition tasks.
%
%
%as suggested  by others \cite{i3d, flowTest, flow2}

\subsubsection*{Inception 3D} 

I3D is created by converting all the filters and pooling kernels in Inception v1 into a 3D CNN. Squared filters of size N $\times$ N are made cubic and becomes N $\times$ N $\times$ N filters. The pre-trained 2D ImageNet weights from Inception v1 are repeated along the  time dimension and rescaled by normalization over N. The 3D version is further trained on the large activity recognition dataset, Kinetics Human Action Video Dataset which has 400 different classes and over 400 clips per class. An I3D model is trained for both data representations, i.e. optical flow and RGB stream. % During testing I3D average the logits outputted from the two networks before applying a softmax to generate the final output.

%
%
%Thus, 
%
%In order to be able to recognize the resuscitation activities a certain video frame rate is necessary to distinguish short sequences of repetitive activities, such as stimulation and ventilation, from random HCPH and BMR movements. Thus, videos of frame rates below 5 frames per second
%
%A linear frame interpolation step is applied to create fixed frame rate videos, suited for usage in activity recognition
%
%
%
%Videos should have a fixed and  reasonable frame rates are ne
%Since the newborn resuscitation videos are of very variable quality and has frame rates ranging from 2-30 frames per second (FPS) some pre-processing steps are performed to ensure quality data. 

\begin{figure*}[h]
\centering
% \hspace*{-1cm} 
\includegraphics[width=\linewidth]{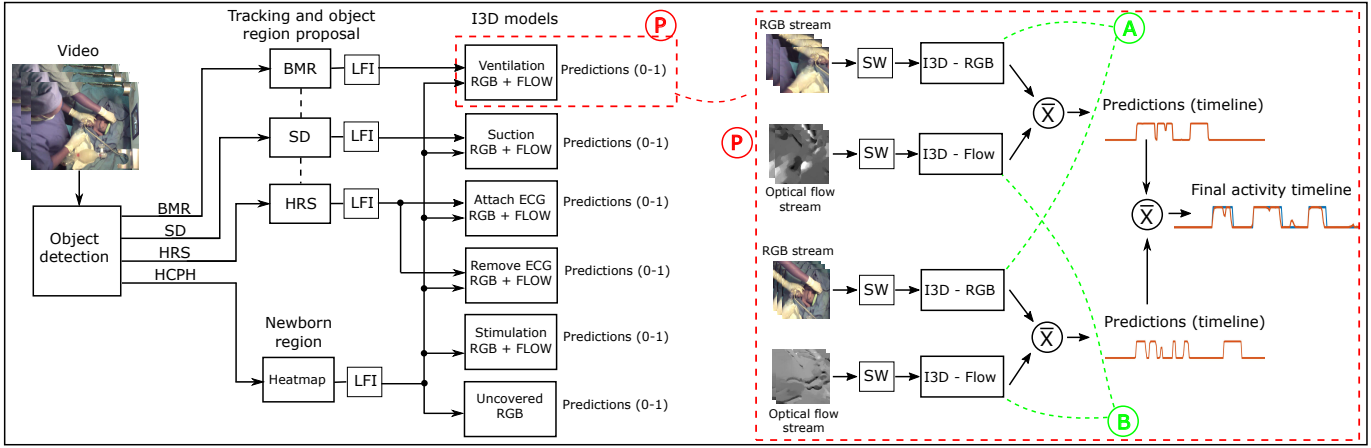}
\caption{An overview of the proposed system for automatic recognition of the newborn resuscitation activities. Depending on the activity, class-relevant regions are analyzed in class-relevant Inception 3D architecture models trained on the specific activity to do binary classification - activity or no activity. The predictions from activities involving two regions and models are averaged before generating the final activity timeline. The SW blocks illustrate that a sliding window of the stream is used as the model`s input.
%
%Activity recognition examples of activities that are object dependent. Both object region and newborn region is input to the same RGB and Flow models trained on RGB and flow data respectively to do binary classification on that specific activity. The predictions from each of the inputs are averaged providing the final activity timeline.
}  
\label{fig:actRecEx}
\end{figure*}

\subsubsection*{Predictions and Timeline generation}
\label{timelinePaper6}

Figure \ref{fig:actRecEx} illustrates how the timelines are predicted and generated for  different activities. The red squared section marked \textit{P} shows in more detail how the prediction of the activity \textit{ventilation} is performed using both the BMR region and the newborn region, and both an RGB model and an optical flow model.
Since the activities can overlap in time and the video quality makes the activity recognition difficult, the models are trained on individual activities to do binary classification - activity or no activity. 
%
% The models are trained on that specific activity to do binary classification - activity or no activity. 
The object dependent activities that analyze both an object region and a newborn region use the same RGB and Flow models in the two analyses, indicated with \textit{A} and \textit{B}, and the models are trained on data from both regions. This is done similarly for all the 6 activities, resulting in 11 different I3D models learned (6 RGB and 5 flow).
From the left in Figure \ref{fig:actRecEx}:  First, a video undergoes object detection, tracking, and region proposal. Next, the videos from the regions are linear frame interpolated and optical flow is estimated using the  TV-L1 algorithm \cite{flow}, as proposed by \cite{i3d}. Further, a sliding window (SW) generates sub-signals of the RGB stream and the optical flow stream, and the sub-signals are fed to their corresponding model.
%
%Further, the videos from different regions are LFI and the TV-L1 algorithm \cite{flow}, as proposed by \cite{i3d}, estimate optical flow on the relevant regions for the activity, as can be seen in the examples in section \textit{P} of the figure, and short temporal sequences of  RGB and flow data is input to models trained on their respective data representation. 
The logits from the final I3D layer of the two models are averaged before softmax is applied to perform predictions. The predictions from the two activity-relevant regions are further averaged to generate the final predicted timeline for that specific activity.

The activities \textit{stimulation} and \textit{uncovered}  are not  object-dependent, and only the \textit{newborn region} is used to generate the activity timeline. Since the activity \textit{uncovered} is not motion dependent, the computational demanding TV-L1 flow prediction is not performed for this activity and the predictions are generated by using only the RGB data and model. 

\subsubsection*{Estimation of the Number of Health Care Providers}

The timeline estimation of the number of HCP present in the resuscitation episode, $\mathit{\#HCP(i)_{E}}$, is found by counting the number of detected HCPH for each time index $i$, and is consistent with the method we proposed in \cite{objectresuscitation}. The method only does roughly estimate of the number of HCPs present but makes it possible to recognize if no HCP is present and cases where there are certainly more than one HCP present. % as explained in \cite{objectresuscitation}.

\section{Experiments}
In this section, we present 4 different experiments for the two steps in the proposed ORAA-net, Figure \ref{fig:oversiktPaper6}.  For step 1, we present an object detection experiment, Ex.1, where the 4 different network architectures of Table \ref{tab:table1Paper6} are tested.  The best architecture, RetinaNet, is  further used  in a second experiment, Ex.2, to investigate if object tracking and region proposal is improved compared to our recent work employing the YOLOv3 architecture \cite{objectresuscitation}.  
%Both Ex.1, and  Ex.2 are compared to our recent work employing the YOLOv3 architecture  \cite{objectresuscitation}.  

For step 2, Figure \ref{fig:oversiktPaper6}, which is the main experiments producing  timelines of the resuscitation activities, we evaluate the I3D models trained on the specific activities, in Ex. 3, and investigate if the results could be improved by finding optimal thresholds for the generation of the activity timelines, in Ex. 4.

%
%two experiments are perfomed to evaluate step 2 of Figure, \ref{fig:oversiktPaper6}
%The Activity detection experiment in Section \ref{ex_RPPaper6} is the main experiment producing automated timelines of the resuscitation activities.   

\begin{figure}[h]
\centering
% \hspace*{-1cm} 
\includegraphics[width=\linewidth]{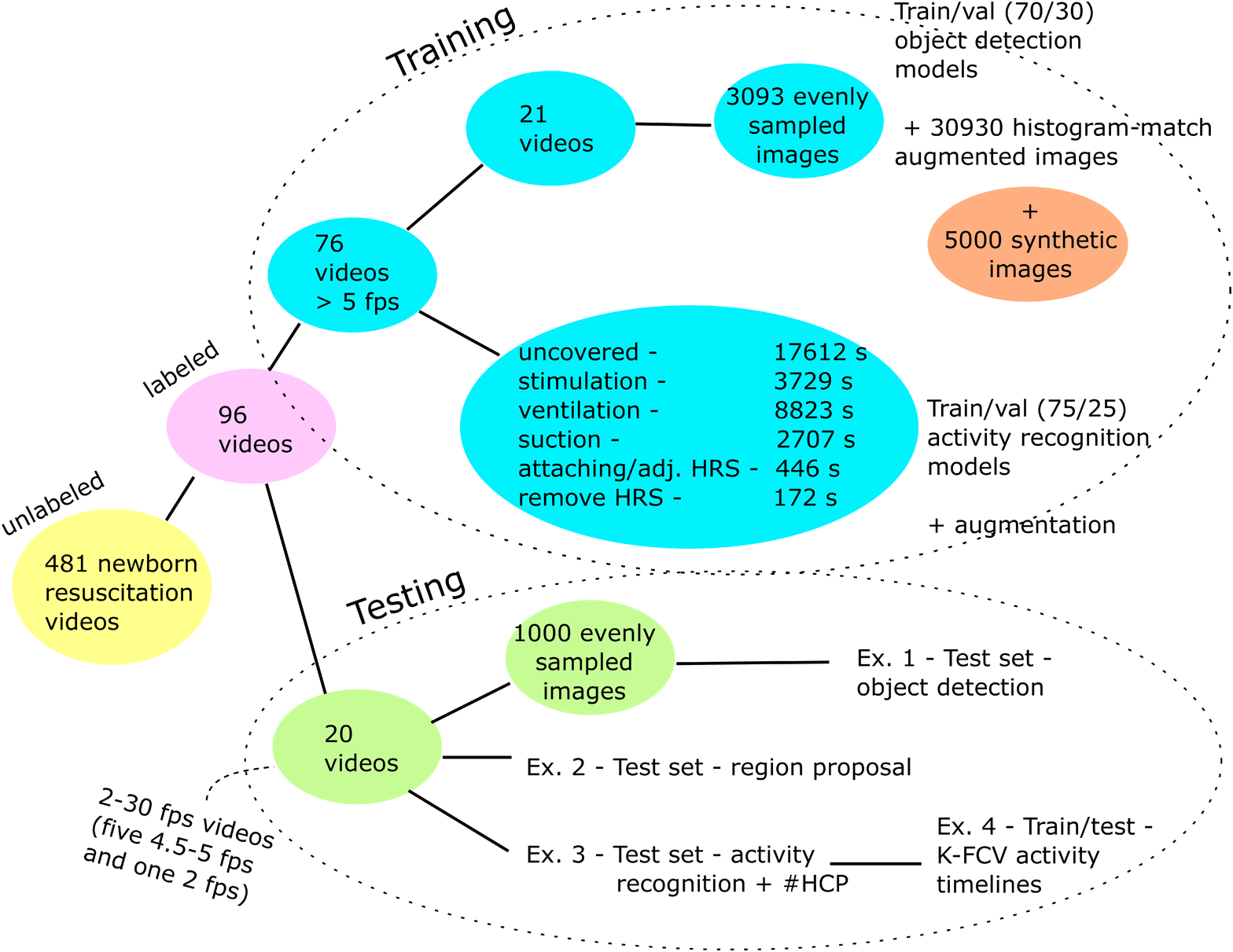}
\caption{Datasets used in training and testing of the 4 experiments. The 76 labeled videos used in training has frame rates $>$ 5 frames per second (fps) and the 20 videos in the test set includes videos with frame rates ranging between 2-30 fps. Except from the 5-fps requirement for the training data, the videos were selected randomly from the 481 newborn resuscitation videos}  
\label{fig:experiments}
\end{figure}

We used the video annotation tool ELAN to manually annotate ground-truth timelines in all videos included in the
training and validation set of Ex. 3, and in the test set for Ex. 2, Ex. 3 and Ex. 4.
%
% test set for Ex. 2 - region proposal, % evaluation of step 1 - object tracking an region proposal 
%and for the training set, validation set, and test set for Ex. 3 - activity recognition. % \ref{fig:oversiktPaper6} and for the training and validation set in part C . 
The following is annotated: 
\begin{itemize}

  \item  Activities
  \begin{itemize} 
      \item  Uncovered: The newborn is not covered by a blanket.
  \item Stimulation: Thoroughly drying and rubbing.
   \item Ventilation: Bag-mask ventilations 
   \item Suction:  Removal of liquid from the mouth/airways.
   \item Attaching/adjusting the ECG sensor
    \item Removing the ECG sensor
    %\item  Chest compressions
\end{itemize}
 \item  The number of HCPs present
\end{itemize}
An overview   of the datasets used in the four experiments can be seen in Figure \ref{fig:experiments}. From the 481 videos in the dataset we use 76 videos, and synthetic and augmented data, split for training and validation of the neural network models. The details for the splits and the different experiments can be seen in Figure \ref{fig:experiments}. In addition, 20 videos are used to create a test set for evaluation of the models. 

%In Table \ref{tab:dataset} the data material used in the 4 experiments are listed.

\begin{table*}[h]
\centering
\caption{Object detection results, measured by mean Average Precision $_{50}$ with associated true positives and false positives. HCPH = health care provider hand, BMR = bag-mask resuscitator, HRS = heart rate sensor and SD = suction device. For the YOLOv3 416 $\times$ 416 network the results presented in \cite{objectresuscitation} is marked with *} 

\begin{tabular}{|l|c|c|c|c|c|c|c|c|c|c|c|c|}
\cline{1-13}
& \multicolumn{3}{|c|}{\textbf{YOLOv3}} & \multicolumn{3}{|c|}{\textbf{RetinaNet}} & \multicolumn{3}{|c|}{\textbf{SSD MultiBox}} & \multicolumn{3}{|c|}{\textbf{Faster R-CNN}} \\ 
& \multicolumn{3}{|c|}{416 $\times$ 416} & \multicolumn{3}{|c|}{1024$\times$ 1280} & \multicolumn{3}{|c|}{512$\times$ 512} & \multicolumn{3}{|c|}{600$\times$ 750} \\ \cline{2-13}
\multicolumn{1}{|c|}{} & AP & TP & FP & AP & TP & FP & AP & TP & FP & AP & TP & FP \\
\cline{1-13}
\multicolumn{1}{|l|}{\textbf{HCPH}} & 71.58 (70.07*) & 1636 (1607*) & 165 (192*) & 74.58 & 1752 & 240 & \textbf{78.34} & 1815 & 231 & 61.85 & 1625 & 507 \\ \hline
\multicolumn{1}{|l|}{\textbf{BMR}} & 65.89 (62.07*) & 644 (602*) & 91 (71*) &  \textbf{69.75} & 708 & 168 & 64.89 & 633 & 85 & 55.14 & 596 & 161 \\ \hline
\multicolumn{1}{|l|}{\textbf{HRS}} &80.44 (79.38*) & 486 (478*) & 32 (22*) & 78.57 & 475 & 58 & \textbf{82.77} & 500 & 33 & 64.18 & 387 & 46 \\ \hline
\multicolumn{1}{|l|}{\textbf{SD}} & 22.98 (19.25*) & 152 (124*) & 156 (89*) & \textbf{44.22} & 265 & 71 & 34.29 & 201 & 16 & 33.23 & 217 & 118 \\ \hline
\multicolumn{13}{|l|}{} \\ \hline
\multicolumn{1}{|l|}{\textbf{mAP$_{50}$}} & \multicolumn{3}{|c|}{60.22 (57.69*)} & \multicolumn{3}{|c|}{\textbf{66.78}} & \multicolumn{3}{|c|}{65.08} & \multicolumn{3}{|c|}{53.60} \\ \hline

\end{tabular}
\label{tab:table2Paper6}
\end{table*}

\subsection{Performance metrics}

In Ex. 1, the object detection
 results were evaluated by use of the Average Precision (AP) and the mean Average Precision (mAP) metrics defined in the PASCAL VOC 2012 challenge. The required accuracy of the localization task was defined by an Intersection over Union (IoU) threshold of 0.5. In addition to the mAP criterion, the number of True Positives (TP) and False Positive (FP) detections were assessed. This was due to FP detections being highly undesirable for object tracking and poor TP/FP ratios not being sufficiently penalized by mAP. As an aid in setting probability thresholds for desirable trade-offs of TPs and FPs, the distribution of probability scores was assessed. Distributions were drawn from network predictions on the validation dataset. %; examples are seen in Figure \ref{fig:dist1}.

In Ex. 2, a performance measure, $P$, from \cite{objectresuscitation}, is used to evaluate the tracking performance of each object-dependent activity and each episode. $P$ is defined by the general equation:
%
%Using the best object detector architecture, RetinaNet, we have evaluated the object detection performance during activity sequences for the objects \textit{SD}, \textit{BMR}, and \textit{HRS}, and compared it to \cite{objectresuscitation}. The test set consists of 20 videos as in \cite{objectresuscitation}. A performance measure, $P$ is used to evaluate the tracking performance of each object dependent activity and each episode, defined by the general equation:

\begin{equation} \label{P}
 P = (\frac{1}{N_s}  \sum_{i=1}^{N_s}{I_f(i)}) *100
\end{equation}
where $N_s$ is the number of frames in the  episode and $I_f(i)$ an indicator function defined as 1 on correct detections if $|detection(i)_{E} - ground truth(i)_{E}| = 0$ and 0 otherwise. 
An object is classified as detected during an activity sequence
if the detection overlaps with the ground-truth data $> 80 \%$ of the
time.

In Ex.3 and Ex. 4 the activity recognition and activity timelines results are evaluated by comparing the ground-truth activity timelines with the predicted activity timelines and estimating True Positivies, (TP), True Negatives (TN), False Positive (FP) and False Negatives (FN). 
To handle class-imbalance in the binary activity classification we evaluate the results by using the two metrics; precision and recall, in addition to the accuracy metric. Ex. 4 also utilize the F1-score in  a K-fold Cross-Validation (K-FCV) experiment.% the and the \textit{accuracy} to add information about the models` ability to predict true negatives (TN)% metrics defined as
%
%The evaluation measurement in Ex 4. is the  \textit{F1-score}, a combination of the \textit{Precision} and \textit{Recall} since both are here considered equally important., 
%%
%%\begin{equation}
%%F1 = 2\frac{Precision \times Recall}{Precision + Recall}
%%\end{equation}
%%The result from the K-FCV experiment, together with the quantile measurement for the threshold used is listed in Table \ref{tab:table5Paper6}. 
%Here, the \textit{accuracy} is calculated to add information about the models` ability to predict true negatives (TN).

\subsection{ORAA-net Step 1 - Object Detection and Region Proposal}

In Ex.1, where we compared different object detectors, RetinaNet and Faster R-CNN employed ResNet-50 \cite{resnet} with pre-trained weights on the ImageNet dataset as the initial network. SSD MultiBox used VGG-16 \cite{vgg} with weights pre-trained on the COCO dataset as the initial network. YOLOv3 was included in the experiment for reference, using the same setup as recently presented by the authors in \cite{objectresuscitation}. The networks were all further trained on a dataset consisting of manually labeled images from the videos, augmented images using histogram matching, and synthetic images. The details can be seen in the upper-right corner of Figure \ref{fig:experiments}.
%
%
%a total of 38095 images, including 2165 labeled images extracted from the video recordings, 30930  images augmented using histogram matching \cite{hist}, and 5000 generated synthetic images. 
The datasets are similar to those used in \cite{objectresuscitation}, however the number of synthetic images were reduced due to experiencing that models were overfitting on synthetic data in the experiments. %Additionally, the previous method of choosing a random 10\% of images from the training dataset for validation was dropped. 
%The validation dataset for the object detection was a selection of 928 labeled images from the video recordings, not a part of the training set. 
A built-in image augmentation pipeline recommended in \cite{ssdmultibox} was used for the SSD MultiBox network, which included random flips, crops and photometric distortions. The object detection networks were tested on the same dataset used in \cite{objectresuscitation}. %, consisting of 1000 labeled images from the video recordings. 

%The results were evaluated by use of the Average Precision (AP) and the mean Average Precision (mAP) metrics defined in the PASCAL VOC 2012 challenge, similar to \cite{objectresuscitation}. The required accuracy of the localization task was defined by an IoU threshold of 0.5. In addition to the mAP criterion, the number of True Positives (TP) and False Positive (FP) detections were assessed. This was due to FP detections being highly undesirable for object tracking and poor TP/FP ratios not being sufficiently penalized by mAP. As an aid in setting probability thresholds for desirable trade-offs of TPs and FPs, the distribution of probability scores was assessed. Distributions were drawn from network predictions on the validation dataset. %; examples are seen in Figure \ref{fig:dist1}.

In Ex. 2, we evaluated the region proposal performance during activity sequences for the objects \textit{SD}, \textit{BMR}, and \textit{HRS} using the best object detector from Ex. 1, RetinaNet, and compared it to \cite{objectresuscitation}.

%($>$ half the object is
%included in the proposed object region)
%In \cite{objectresuscitation} we achieved acceptable detection in 96.97 \% (64/66), 100 \% (43/43) and 75 \% (45/60) of the sequences during \textit{ventilation}, \textit{Attach/remove HRS} and \textit{suction} respectively. With RetinaNet as the new object detector the results where XX.XX \% (64/66), XX.XX \% (43/43) and XX.XX \% (45/60).

\subsection{ORAA-net Step 2 - Activity Recognition and Activity Timelines}

In Ex. 3, the I3D flow and RGB weights pre-trained on ImageNet, and Kinetics 400 \cite{i3d} are further trained on the individual activities to do binary classification - activity or no activity. The threshold, $T_{act}$ for detection of class or not, is here set to 0.5. 
The videos are Linear Frame Interpolated (LFI)   to a fixed frame rate of 15 fps, which are reasonably close to the pre-trained I3D weights that are trained on 25 fps videos. 76 videos are used in training of the models, and the details can be seen in Figure \ref{fig:experiments}.
%
%As shown in Figure \ref{fig:experiments}, 
%76 videos are LFI to a fixed frame rate of 15 fps and are  manually annotated using ELAN to generate training data. 15 fps is reasonably close to the pre-trained I3D weights that are trained on 25 fps videos. %The annotations cover the activities \textit{uncovered}, \textit{stimulation}, \textit{ventilation}, \textit{suction}, \textit{attaching/adjusting ECG} and \textit{removing ECG}, and 
%The total length of the activities \textit{uncovered}, \textit{stimulation}, \textit{ventilation}, \textit{suction}, \textit{attaching/adjusting ECG}, and \textit{removing ECG} from the 76 videos are 17612, 3729, 8823, 2707, 446, and 172 seconds respectively. 75 \% of the training data is used to create a training set and the rest to create a validation set. 

The test set consists of 20 videos as in \cite{objectresuscitation}  and in Ex. 1 and Ex. 2. The test set includes 5 videos of frame rates between 4.5-5 fps and one with a frame rate as low as 2 fps.

The input sequences to the RGB and flow I3D models during training and testing are 45 frames long, corresponding to 3 seconds of activity. The frame size is $256$  x $256$ pixels. During training the activity sequence examples overlap with 1/2, and during testing they overlap with 2/3 - resulting in a new analysis every second.
The training examples are also augmented by random cropping, flipping, 90-degree rotation, noise adding and motion blurring. 
During training we use a batch size of 6 and train for 15000 steps. The learning rate is initially set to 0.0001, and every 3000 step it is decreased  by a factor of 0.1

The experiment also compare the usage of both RGB and Flow models in the predictions versus using the individual models alone.
%In addition to providing the results from predictions generated by combining the RGB and flow models, we also provide the results for the individual models. 
This is performed to investigate if acceptable results can be achieved using only the RGB models since the TV-L1 algorithm is highly computational demanding and limits the possibility for real-time usage.

For the activity \textit{uncovered}, we only use the RGB data and model since this activity is not motion dependent and does not require a motion analysis.

In Ex. 3
 we also evaluate the performance of the estimation of number of HCP present in the videos, and compare it to \cite{objectresuscitation}.

%\begin{equation}
%Precision = \frac{TP}{TP + FP}, \quad \quad Recall = \frac{TP}{TP + FN}
%\end{equation}
%
%
%

In Ex. 4, a K-fold Cross-Validation (K-FCV) experiment is performed to find the optimal $T_{act}$ for the best I3D models from Table \ref{tab:table4Paper6} for each of the included activity classes. $K$ is set to 20, i.e., one fold for each of the 20 videos included in the test set. 
%The evaluation measurement in the experiment is the  \textit{F1-score}, a combination of the \textit{Precision} and \textit{Recall} since both are here considered equally important., 
%%
%%\begin{equation}
%%F1 = 2\frac{Precision \times Recall}{Precision + Recall}
%%\end{equation}
%The result from the K-FCV experiment, together with the quantile measurement for the threshold used is listed in Table \ref{tab:table5Paper6}. Here, the \textit{accuracy} is calculated to add information about the models` ability to predict true negatives (TN). %\textit{Accuracy} is defined as
%\begin{equation}
%\mathit{Accuracy= \frac{TP+TN}{TP+TN+FP+FN}}
%\end{equation}

\section{Results}

\subsection{ORAA-net Step 1 - Object Detection and Region Proposal}
\label{results}

%\begin{figure}[h]
%\centering
%% \hspace*{-1cm} 
%\includegraphics[width=\linewidth]{Figures/TP-FP.eps}
%\caption{Distribution example of true and false positives in object detection results. The example is from RetinaNet detections of the bag-mask resuscitator (BMR) on the validation dataset.The red line represents the chosen threshold for this specific network and class.}  
%\label{fig:dist1}
%\end{figure}

The results for Ex. 1 are listed in Table \ref{tab:table2Paper6}. Both the RetinaNet architecture and the SSD MultiBox architecture show improved object detection results compared to the YOLOv3 416 $\times $ 416 architecture used in \cite{objectresuscitation}, with RetinaNet as the overall winner. In \cite{objectresuscitation} we also split the original images into five 608 $\times $ 606 sub-images and achieved an mAP close to the one achieved by RetinaNet for the class SD, mAP 42.02, but with a much smaller TP/FP ratio, YOLOv3`s 1.316 vs. RetinaNet`s 3.73.

The results for Ex. 2 are listed in Table \ref{tab:table3Paper6}. Here, the best object detector architecture, RetinaNet, resulted in a large improvement in the detection of the SD during the \emph{suction} activity.

%We also evaluated the performance of the predicted number of HCP present during the resuscitation episodes.  All results can be seen in Table \ref{tab:table3Paper6}.

\begin{table}[h]
\caption{Performance results for the object detection when relevant activities occurs  - using the RetinaNet architecture \cite{retinanet} (\# detected / \# true). The results presented in \cite{objectresuscitation} are marked with *. } 
\centering
\begin{tabular}{lcc}

\cline{1-3}
\hline
\multicolumn{1}{|l|}{\textit{\textbf{\begin{tabular}[c]{@{}l@{}}Object detection \\ during activity\end{tabular}}}} & \multicolumn{1}{c|}{\textbf{$P$}} & \multicolumn{1}{c|}{\textbf{Activities}} \\ \hline
\multicolumn{1}{|l|}{\textbf{BMR}} 	& \multicolumn{1}{c|}{96.97 \%  (96.97*)} 			& \multicolumn{1}{c|}{Ventilation} \\ \hline
\multicolumn{1}{|l|}{\textbf{\begin{tabular}[c]{@{}l@{}}HRS\end{tabular}}} 	& \multicolumn{1}{c|}{100 \% (100*)} & \multicolumn{1}{c|} {Attach/remove HRS} \\ \hline
\multicolumn{1}{|l|}{\textbf{SD}} 		& \multicolumn{1}{c|}{88.33 \% (75.00*)} 		& \multicolumn{1}{c|}{Suction} \\ \hline

\end{tabular}
\label{tab:table3Paper6}
\end{table}

\subsection{ORAA-net Step 2 - Activity Recognition and Activity Timelines}

Table \ref{tab:table4Paper6} shows the results for Ex.3, activity recognition using the I3D models for the activities \textit{uncovered}, \textit{stimulation}, \textit{ventilation}, \textit{suction}, \textit{attach/adjust HRS}, and \textit{remove HRS}.
%For Ex. 3, the detection results for the activities \textit{uncovered}, \textit{stimulation}, \textit{ventilation}, and \textit{suction}, and the the models, RGB, Flow, and RGB+Flow can be seen in Table \ref{tab:table4Paper6}. 
In Table \ref{tab:numHCP},  the results from the estimation of the number of HCP present in the resuscitation episodes are presented. Here, the results are compared to the results from \cite{objectresuscitation}.

\begin{table}[h]
\caption{detection results for the activities \textit{uncovered}, \textit{stimulation}, \textit{ventilation}, \textit{suction}, \textit{attach/adjust Heart Rate Sensor (HRS)}, and \textit{remove HRS} using the the models I3D-RGB, I3D-Flow, and I3D-RGB+Flow}
\begin{tabular}{|c|c|c|c|c|c|c|}
\hline
\multirow{2}{*}{}                                                              & \multicolumn{2}{c|}{\textbf{\scriptsize I3D \footnotesize RGB}} & \multicolumn{2}{c|}{\textbf{\scriptsize I3D \footnotesize Flow}} & \multicolumn{2}{c|}{\textbf{\scriptsize I3D \footnotesize RGB+Flow}} \\ \cline{2-7} 
                                                                               & \textbf{Prec.}  & \textbf{Rec.} & \textbf{Prec.}  & \textbf{Rec.}  & \textbf{Prec.}    & \textbf{Rec.}   \\ \hline
\textbf{\begin{tabular}[c]{@{}c@{}}Uncovered\end{tabular}}                 & 92.89           & 78.74           & -               & -                & -                  & -                 \\ \hline
\textbf{\begin{tabular}[c]{@{}c@{}}Stimulation\end{tabular}}               & 75.48           & 73.13           & 67.45           & 79.80            & 79.41              & 78.15             \\ \hline
\textbf{\begin{tabular}[c]{@{}c@{}}Ventilation\end{tabular}}               & 81.70           & 83.57           & 80.03           & 84.26            & 88.64              & 88.34             \\ \hline
\textbf{\begin{tabular}[c]{@{}c@{}}Suction\end{tabular}}                   & 44.96          & 59.81
 & 44.98          & 64.92            & 56.01              & 65.61            \\ \hline

 \textbf{\begin{tabular}[c]{@{}c@{}}Att./adj. HRS\end{tabular}}                   & 56.02  & 49.18   & 23.84 & 45.88    & 50.00  & 50.83           \\ \hline

\textbf{\begin{tabular}[c]{@{}c@{}}Remove HRS\end{tabular}}                   & 4.59  & 58.49   & 3.06 & 45.28    & 6.73    & 52.83           \\ \hline

%\textbf{\begin{tabular}[c]{@{}c@{}}Removing \\ ECG\end{tabular}}               & 4.29            & 61.70           & 3.06            & 51.06            & 6.49               & 57.45             \\ \hline
%\textbf{\begin{tabular}[c]{@{}c@{}}Attach/adjust\\ ECG\end{tabular}} & 55.26           & 50.52           & 23.84           & 47.77            & 50.00             & 52.92              \\ \hline
\end{tabular}
\label{tab:table4Paper6}
\end{table}

\begin{table}[h]
\caption{Performance results of the prediction of the number of health care providers (HCPs) - using the RetinaNet architecture \cite{retinanet}. The results presented in \cite{objectresuscitation} are marked with *. Q denotes quartile measurements.}  
\centering
\begin{tabular}{lcc}

\cline{1-3}
\hline

\multicolumn{1}{|l|}{\textit{\textbf{\begin{tabular}[c]{@{}l@{}}HCP detection \end{tabular}}}} & \multicolumn{1}{c|}{\textbf{$\overline{P}$}} & \multicolumn{1}{c|}{\textbf{Q  (25,50,75)}}\\ 
\hline
%\multicolumn{1}{|l|}{\textbf{No HCP}} 		& \multicolumn{1}{c|}{72.06 \% (90.70*) } 				& \multicolumn{1}{c|}{} \\ \hline
%\multicolumn{1}{|l|}{\textbf{One HCP}} 		& \multicolumn{1}{c|}{ 72.30 \% (90.48*)} 				& \multicolumn{1}{c|}{} \\ \hline
%\multicolumn{1}{|l|}{\textbf{Two HCPs}} 	& \multicolumn{1}{c|}{67.78 \% (53.31*)} 				& \multicolumn{1}{c|}{} \\ \hline
%\multicolumn{1}{|l|}{\textbf{\begin{tabular}[c]{@{}l@{}}Three (or more)\\ HCPs\end{tabular}}} & \multicolumn{1}{c|}{ 23.20 \% (6.88*) } & \multicolumn{1}{c|}{} \\ \hline
%
%\multicolumn{1}{|l|}{\textit{\textbf{\begin{tabular}[c]{@{}l@{}} \end{tabular}}}} & \multicolumn{1}{c|}{\textbf{{$\overline{P}$}}} & \multicolumn{1}{c|}{\textbf{Q  (25,50,75)}} \\ \hline
\multicolumn{1}{|l|}{\textbf{\begin{tabular}[c]{@{}l@{}}HCP correct pred.\end{tabular}}} & \multicolumn{1}{c|}{68.32 \% (71.16*)} & \multicolumn{1}{c|}{\begin{tabular}[c]{@{}c@{}}53.86, 75.64, 85,46\\ (50.72, 78.56, 89.45 *) (\%)\end{tabular}} \\ \hline

\multicolumn{1}{|l|}{\textit{\textbf{\begin{tabular}[c]{@{}l@{}} \end{tabular}}}} & \multicolumn{1}{c|}{\textbf{{$\overline{E}$}}} & \multicolumn{1}{c|}{\textbf{}} \\ \hline
\multicolumn{1}{|l|}{\textbf{\begin{tabular}[c]{@{}l@{}}HCP  pred. error\end{tabular}}} & \multicolumn{1}{c|}{0.34 (0.32*)} & \multicolumn{1}{c|}{\begin{tabular}[c]{@{}c@{}}0.15 0.25, 0.48\\ (0.11    0.22    0.54 *)\end{tabular}} 
 \\ \hline
\end{tabular}
\label{tab:numHCP}
\end{table}

\begin{figure}[h]
\centering
\includegraphics[width=\linewidth]{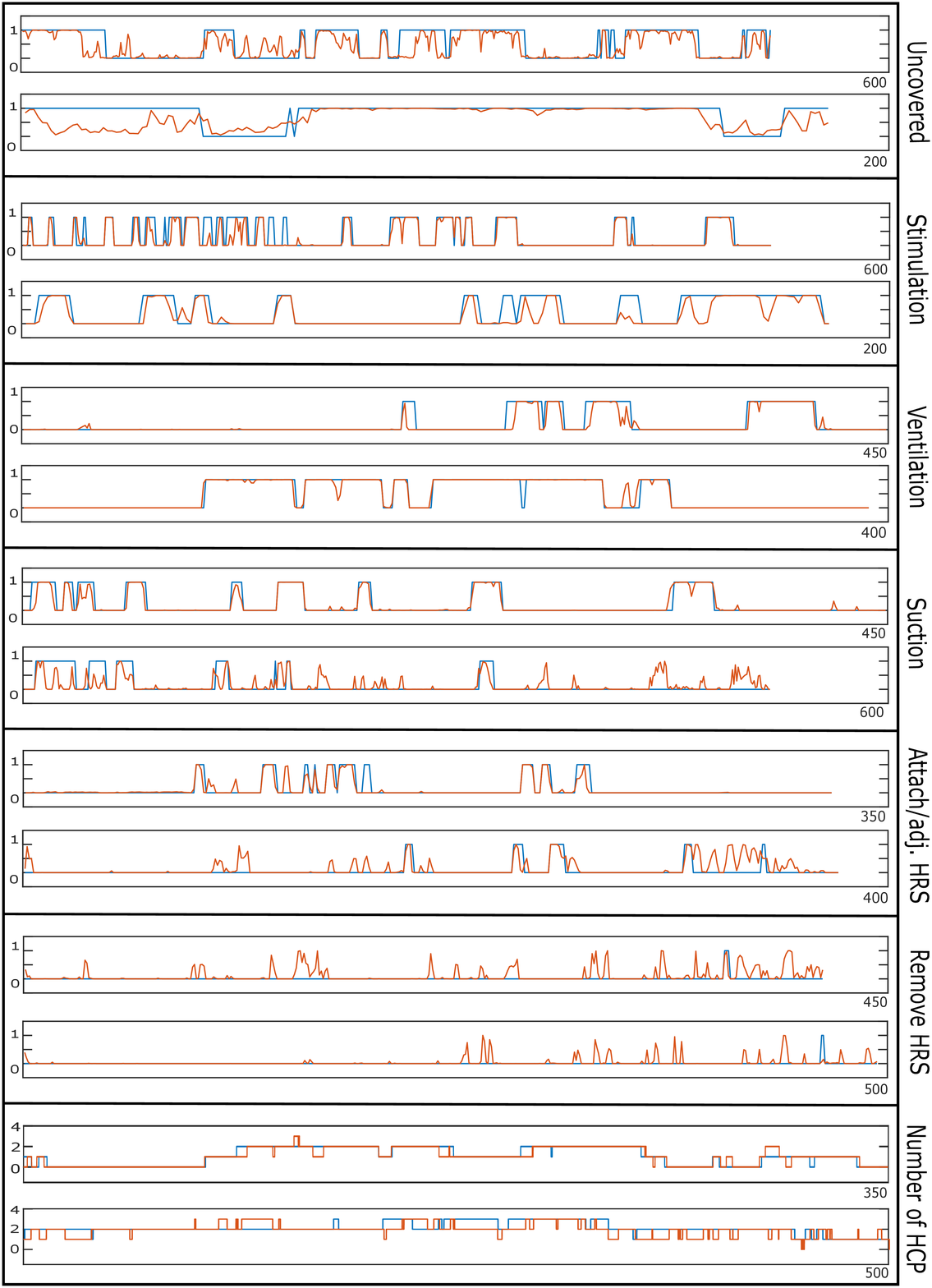}
\caption{Examples of activity detection results for the activities  \textit{Uncovered}, \textit{Stimulation},  \textit{Ventilation}, \textit{Suction}, \textit{Attach/adjust Heart Rate Sensor (HRS)}, and \textit{Remove HRS}, and  the \emph{Number of health care providers (HCP)} estimated from the detected HCP`s hands. Two test set examples that illustrate both strengths and weaknesses are chosen for each activity. The y-axis represent the probability for the activity, between 0 and 1, and x-axis the video length in seconds. Blue lines represent the reference data from the manual annotations and orange lines  the detection results.}
\label{fig:ventAndSuct}
\end{figure}

Figure \ref{fig:ventAndSuct} shows two test set video examples of the raw output for the timeline predictions of the 6 activity detected by the I3D models, and two examples of the estimated number of HCP present in the resuscitation. The predictions are shown with orange lines, and the reference data with blue lines.

Table \ref{tab:table5Paper6} shows the results for Ex. 4, the K-fold cross validation experiment. The Table lists both the mean results for the four therapeutic activities, and the overall mean results for all the six activities.

%The results for \textit{Attach/adjust HRS} was a \textit{precision} of 50 \% and a \textit{recall of 52.92}, and the results for \textit{Removing HRS} was a \textit{precision} of 6.49 \% and a \textit{recall} of 57.45. 
%Examples from results for five of the videos in the test set are shown in the upper part  of Figure \ref{fig:ventAndSuct}.

\begin{table}[h]
\centering
\caption{K-fold cross validation threshold test  for activity recognition using the combination of Inception 3D models that provided the best results in Table \ref{tab:table4Paper6}. }
\begin{tabular}{|c|c|c|c|c|c|}
\hline
\multicolumn{1}{|l|}{} & \multicolumn{5}{c|}{\textbf{Activity recognition - I3D}}                                                                                               \\ \hline
                       & \textbf{Mod.} & \multicolumn{4}{c|}{\textbf{K-FCV threshold test}}                                                                   \\ \hline
                       & \textbf{}     & \textbf{Prec.} & \textbf{Rec.} & \textbf{Acc.} & \textbf{\begin{tabular}[c]{@{}c@{}}Thresh., Q 
                       \\ (25, 50, 70)\end{tabular}} \\ \hline
\textbf{Uncovered}     & \scriptsize RGB           & 87.75              & 83.99           & 88.31             & .29, .29, .29                                                            \\ \hline
\textbf{Stimulation}   & \scriptsize RGB+Flow      & 78.79              & 74.59           & 91.61             & .46, .50, .80                                                            \\ \hline
\textbf{Ventilation}   & \scriptsize RGB+Flow      & 87.30              & 90.64           & 96.90             & .34, .34, .34                                                            \\ \hline
\textbf{Suction}       & \scriptsize RGB+Flow      & 56.85              & 61.32           & 92.78             & .51, .51, .51                                                            \\ \hline
\textbf{}       &     &             &        &          &                                                           \\ \hline
\textbf{{\begin{tabular}[c]{@{}c@{}}mean \\ therapeutic \\activities \end{tabular}}}       &     & \textbf{77.67}             & \textbf{77.64}           & \textbf{92.40}            &                                                           \\ \hline

\textbf{}       &     &             &        &          &                                                           \\ \hline

\textbf{Att./adj. \scriptsize HRS}       & \scriptsize RGB+Flow      & 52.65            & 47.77          & 96.76             & .51, .60, .60                                                           \\ \hline

\textbf{Remove \scriptsize HRS}       & \scriptsize RGB+Flow      & 10.24              & 27.67          & 98.27           & .86, .92, .92\\ \hline

\textbf{}       &     &             &        &          &                                                           \\ \hline
\textbf{mean all}       &     & \textbf{62.26}             & \textbf{64.00}           & \textbf{94.11}            &                                                           \\ \hline

\end{tabular}
\label{tab:table5Paper6}
\end{table}

\section{Discussion}
\label{sec:discussionPaper6}

%This Section presents the discussion for the experiments and results presented in Section \ref{sec:exPaper6}.

\subsection{ORAA-net Step 1 - Object Detection and Region Proposal}
\label{d_ODPaper6}

The results give reason to suggest the  RetinaNet as the overall best architecture for this specific task - object detection during noisy newborn resuscitation videos. 
Compared to the YOLO v3 architecture and our results in  \cite{objectresuscitation}, the RetinaNet architecture gave a large increase in AP for SD-detection with an acceptable number of false positives. 

The comparison of the networks in Table \ref{tab:table1Paper6} indicate that using a larger selection of feature map scales was crucial for the improvement. Producing predictions from different sized scales allow the networks to easier recognize objects of both small and large sizes. 
Tsung-Yi Lin et al. also emphasize that RetinaNet are capable of state-of-the-art results due to their novel focal loss \cite{retinanet}.  
The focal loss function introduces a weight 
term that down-weights easy training examples, i.e. examples where the predicted confidence score is high, during training. Thus, the main contributions in the estimated loss come from predictions with low confidence score, and the network is better equipped to handle the class imbalance between background/negatives and objects.

%The focal loss down-weighs easy training examples, i.e where the prediction probability is high, and force the network to focus on difficult examples during training and weight-update. Thus, the focal loss allows RetinaNet to between background and object classes one-stage detectors often str
%
%In one-stage detectors such as Yolo v3 and RetinaNe t this

 %rather than innovations in network design... % "We emphasize that our simple detector achieves top results not based on innovations in network design but due to our novel loss." fra artikkelen. Kan muligens brukes som et argument? 
%SSD MultiBox also utilizes multiple feature map scales for detections, however it does not employ an in-network FPN such as RetinaNet or YOLOv3, but instead use feature maps produced from supplementary convolutional layers following the base CNN. % Det kan muligens argumentere for at SSD ikke gjør det like sterkt som RetinaNet for SP? 

% Bildeoppløsning? Dette er helt sikkert noe som burde blitt sett mer på av meg i masteren, men ble i stedet valg litt ut ifra standard settings.. Men fult mulig at dette er et poeng som kan diskuteres rundt i og med at retinanet skiller seg en del ut.. Har lest en plass at dypere CNNs (som resnet) krever større oppløsning, men finner ikke tilbake til det..

%number of hands - poor results - overall poorer, still struggles with no gloves, but also now a bit more false detections.
%\subsection{Region Proposal - Object Tracking  - Discussion}
%\label{d_RPPaper6}

Using RetinaNet as the base for our object detector resulted in a substantial improvement in the detection of the SD during activity compared to what we achieved in \cite{objectresuscitation} - the performance increased from 75 \% to 88.33 \%. 
%As a consequence, the tracking becomes much more co much smoother tracking of the object. 

Although the object detector benefits from using histogram match augmentation and synthetic images in the training, we will consider pre-processing steps that could standardize the images in future work instead of attempting to create all the variations by augmenting the training data.

\subsection{ORAA-net Step 2 - Activity Recognition  and Activity Timelines}
\label{d_RPPaper6}

The results for the activities 
presented in Table \ref{tab:table4Paper6} and Table \ref{tab:table5Paper6} demonstrate that activity recognition from noisy low-quality videos recorded %from videos recorded 
during newborn resuscitation %in these noisy low quality newborn resuscitation videos where sometimes the activities are largely occluded, 
could be achieved using the presented pre-processing steps for region proposal and the I3D network architecture for temporal analysis.
%as the base models for our activity recognition models - 
%even 
%on noisy low quality videos %where noisy
%%low quality and sometimes the
%activities are largely occluded. 
The results also show that although the TV-L1 optical flow algorithm is highly computational demanding and thus limit the possibilities for real-time usage, we achieve better performance  by using both RGB and optical flow data representations when predicting the activities, as suggested  by others \cite{i3d, flowTest, flow2}.

A larger ablation study should also be performed to identify where the system could be further improved. Since the inputs to the activity recognition models in the ORAA-net step 2 are based on analyzing regions outputted by the ORAA-net step 1, it would be interesting to evaluate if we would achieve improved activity recognition results if the inputs were regions from perfect object detection and tracking. For this to be carried out we would need to label all objects used to propose the regions in each frame of the 20-video test set, which would be very time consuming.

For the activity \textit{uncovered} most of the examples where the models failed to recognize the activity is in cases where the newborn is only partially covered, and the ground truth can be a matter of definition.

For the activity \textit{stimulation}, where we labeled massaging and both large and small stimulation sequences as \textit{stimulation},  the models sometimes struggle to identify the sequences. Especially for cases where the \textit{stimulation} is performed under the newborn lying on its back. Here, we also experienced more false detections in the videos of lower frame rates, and by excluding the 6 videos that originally had 5 fps or lower from the 20 video test set, 
%
%For the activity \textit{stimulation} we also achieved better result by removing  the 6 videos that was recorded with a frame rate of 5 fps or lower from the 20 video test set. Here, 
the precision increased from 79.41 to  82.36 \% and the recall from 78.15 to 81.86 \%. This supports the problem explained in Section \ref{dataGenPaper6} - that low frame rates make recognition of activities that involve fast and large movement more difficult.
From Table \ref{tab:table5Paper6} we can also see that different from the other activities, the quantile measurement for the thresholds used in the \textit{stimulation} K-FCV test is highly dependent on which video is used in the validation, causing both the precision and  the recall to drop compared to when 0.5 was used as the threshold in table \ref{tab:table4Paper6}.

For the activity \textit{ventilation}, the results is highly promising, with a precision and a recall of almost 90 \%.

The models have difficulties with the activity \textit{suction}, where the precision and recall are around 60 \%. 
Examples of two detection results can be seen in the middle section of Figure \ref{fig:ventAndSuct}. The first example demonstrate the models` ability to recognize the activity, and the second one illustrate that the models suffer from both FP and FN.
%The two timelines in the middle of Figure \ref{fig:ventAndSuct} show examples from five of the videos. The first, third, and partially the fourth example shows that the models can output relatively accurate predictions, but the second and fifth illustrate that the models suffer from both FP and FN. 
When considering that the object detector had difficulties recognizing the SD object itself, it is reasonable to believe that this could also be the case for the I3D models. Thus, an explanation could be that the activity`s movement, where the SD is moved back and forth between the newborn`s mouth or nose and the resuscitation table, are sometimes not distinguishable enough from other hand movements occurring during the resuscitations. %More training data could perhaps reduce this problem. 
This is also the activity out of the presented four therapeutic activities in Table \ref{tab:table4Paper6} with the smallest amount of training data - 2707 seconds of unaugmented \textit{suction} activity (see Figure \ref{fig:experiments}).
 %
 %When considering the fact that the object detector struggled with recognizing the SP object itself, it is reasonable to believe that this could also be the case for the I3D models.  
%
%
% the models could experience the activity very similar to other movements. 
The videos with the poorest results were videos of poor quality, i.e. motion-blurred and unfocused, and videos where the camera are positioned further away from the resuscitation table relative to other videos. It also had more difficulties with videos recorded with a wide-angle format, suggesting the need for a pre-processing step to standardize the videos in some way. Excluding videos of poor quality would most likely increase the performance of the activity recognition, but considering that this would remove most of the videos in the present dataset, this option would limit our data analysis.

%In addition to the four activities discussed, we also attempted to detect when the HRS was attached to the newborn. Since the newborn is covered most of the time, thus causing the HRS to not being visible, we tried to recognize when the HRS was attached and when it was removed. In addition, we wanted to see if we could detect when it`s position on the newborn was being adjusted. We generated training examples for one activity class where we included \textit{attaching} or \textit{adjusting} HRS and another class for \textit{removing} HRS. They where each trained on separate models and the results for \textit{Attach/adjust HRS} was a \textit{precision} of 50 \% and a \textit{recall of 52.92}, and for \textit{Removing HRS} the results was a \textit{precision} of 6.49 \% and a \textit{recall} of 57.45.
%
The results for the classes  \textit{Attach/adjust HRS} and  \textit{Removing HRS} are poor. 
As illustrated in Figure \ref{fig:experiments}, the occurrences of these two activities in the 76 videos used in training, was relatively short, 446, and 172 seconds, and considering that a deep neural network  requires a lot of training data, the results are  understandable. However, we did observe that the models learned to recognize the activities in some cases, as can be seen in Figure \ref{fig:ventAndSuct}. % and expect that by including more training data we could further improve these results. .

In the estimation of the number of HCP present we achieved slightly poorer results using the RetinaNet architecture than with the YOLO v3 architecture \cite{objectresuscitation}. As before, the network struggled to recognize hands in poor video quality due to motion smoothing, but also in cases where the HCP does not wear gloves. % The latter could most likely be improved by generating and adding more training examples. 
The proposed method is based on counting the number of detected HCP hands in each frame, which is a quite naive approach that require all hands to be visible at all time. 
Moreover, the method only does roughly estimate of the number of HCPs present and a detection of four HCPHs could correspond to 2-4 HCPs present. 
However, the method makes it possible to recognize if no HCP is present and cases where there are certainly more than one HCP present. A better approach would most likely be to detect both right and left hands, but with these low-quality videos it is very difficult to discriminate between the two.  Moreover, in some videos the camera is positioned in a side-position, causing HCPs to occlude other HCP’s hands. A potential solution for future recordings could be to include an additional camera, positioned further away, that could be used to recognize the number of HCPs participating in the resuscitation.

%The proposed method is based on counting the number of detected HCPH
%in each frame, which is a quite naive approach that require all hands
%to be visible at all time. This is often not the case in these
%videos where the camera could be placed in a side position, causing the
%HCPs to occlude other HCP hands. A better
%approach would most likely be to detect both right and left hands, but
%with these low-quality videos it is very difficult to discriminate between the
%two. A potential solution for future recordings could be to include an additional camera, positioned further away, that could be used to recognize the number of HCPs participating in the resuscitation. 

%In future work a possible solution could be to include an extra camera that records 

In \cite{objectresuscitation}, we also suggested that it would be possible to recognize \textit{chest compressions}, but because  this activity only occurred in 2 of the 76 videos annotated for training, and none of the videos in the test set, we made no attempt of training I3D models for this activity. This activity could instead, as suggested by Vu \cite{huyen} and Gonzalez-Otero \cite{gonzalez2015chest}, be detected from the ECG signal measured with the HRS.

Some of the discussed problems should be possible to solve by further training of the models on more data. This would especially be true for the activities with the smallest amount of training data. Manual annotations are expensive and time consuming, and a potential solution could be to record resuscitation simulation on a manikin, where we focus on the cases where the models struggle to recognize the activities.

The results suggest that we should consider pre-processing steps that could simplify our data in future work. The videos contain a lot of variations and by standardizing them in some way it could be easier for the I3D models to learn the relevant features.

There are also some video quality limitations in the current dataset, e.g. motion blurring, low frame rates etc., making it challenging to recognize activities - regardless of the amount of training data or the deep learning architecture used. In future recordings this could be solved by clearly defining a protocol for standardization of video recordings for automatic activity recognition in a newborn resuscitation setting.

\section{Conclusion and Future Work}
\label{sec:conclusionPaper6}

The results suggest that the proposed two-step ORAA-net, utilizing object detection and tracking to propose detection regions for temporal activity analysis, is well suited for activity recognition  in  noisy and low-quality newborn resuscitation videos where sometimes the activities are largely occluded.

Potential future applications for such a system could be to implement it on-sight, as a real-time feedback system, or as a debriefing tool for newborn resuscitation training. The ORAA-net could  also be adapted for in-hospital emergency situations involving adults.

In future work, we will investigate if adding more training data to the object detection and the activity recognition could further improve the system and make it possible to detect if the HRS is attached to the newborn or not. 
%In addition to investigating the effect of including the optical flow models during prediction, we plan to do an ablation study to evaluate each of the two steps in the ORAA-net. This would allow us to identify which part of the system that is most critical for further improvement.  
We also plan an ablation study to identify where the system could be further improved.
The system could potentially also be simplified by using fewer models in the activity recognition. A possible solution could be to train activities that cannot overlap in time, such as \emph{ventilation} and  \emph{suction}, in the same model.
%In addition we plan to experiment with different  network structures in the activity recognition, 
Besides, we plan to develop a multisensor fusion system that  incorporate the activity recognition from the ECG signals \cite{huyen} into our video-based method in an attempt to increase the system`s performance.

\section{Acknowledgement}

\subsection{Funding}

Our research is part of the Safer Births project which has received funding from: Laerdal Global Health, Laerdal Medical, University of Stavanger, Helse Stavanger HF, Haydom Lutheran Hospital, Laerdal Foundation, University of Oslo, University of Bergen, University of Dublin – Trinity College, Weill Cornell Medicine and Muhimbili National Hospital. The work was partly supported by the Research Council of Norway through the Global Health and Vaccination Programme (GLOBVAC) project number 228203.

For the specific study of this paper; Laerdal Medical provided the video equipment. Laerdal Global Health funded data collection in Tanzania and IT infrastructure. The University of Stavanger funded the interpretation of the data.

\subsection{Ethical approval}

This study was approved by the National Institute of Medical Research (NIMR) in Tanzania (NIMR/HQ/R.8a/Vol. IX/1434) and the Regional Committee for Medical and Health Research Ethics (REK), Norway  (2013/110/REK vest). All women were informed about the ongoing studies, but consent was not deemed necessary by the ethical committees for this descriptive sub-study.

\subsection{Conflict of interests}

Myklebust is employed by Laerdal Medical. He contributed to study design and critical revision of the manuscript, but not in the analysis and interpretation of the data.

\color{black}
\bibliographystyle{IEEEtran}
\bibliography{ActivityRecognition_refs}
\end{document}